\documentclass{iopconfser}
\usepackage{mathtools}
\usepackage{amsfonts}
\usepackage{subcaption}
\usepackage[hidelinks]{hyperref}

\DeclareMathOperator*{\argmax}{arg\,max}
\renewcommand{\vec}{\boldsymbol}

\begin{document}

\title{Simulation-based Inference of Massive Black Hole Binaries using Sequential Neural Likelihood}

\author{Iván Martín Vílchez$^{1,2}$, Carlos F. Sopuerta$^{1,2}$}

\affil{$^1$Institut de Ciències de l’Espai (ICE, CSIC), Campus UAB, Carrer de Can Magrans s/n, Cerdanyola del Vallès 08193, Spain}
\vspace{0.4cm}
\affil{$^2$Institut d’Estudis Espacials de Catalunya (IEEC), Edifici RDIT, C/ Esteve Terradas, 1,
desp. 212, Castelldefels 08860, Spain}

\email{imartin@ice.csic.es}

\begin{abstract}
We propose a machine learning-based approach for parameter estimation of Massive Black Hole Binaries (MBHBs), leveraging normalizing flows to approximate the likelihood function.
By training these flows on simulated data, we can generate posterior samples via Markov Chain Monte Carlo with a relatively reduced computational cost.
Our method enables iterative refinement of smaller models targeting specific MBHB events, with significantly fewer waveform template evaluations.
However, dimensionality reduction is crucial to make the method computationally feasible: it dictates both the quality and time efficiency of the method.
We present initial results for a single MBHB with Gaussian noise and aim to extend our work to increasingly realistic scenarios, including waveforms with higher modes, non-stationary noise, glitches, and data gaps.
\end{abstract}

\section{Introduction} 
Massive Black Hole Binaries (MBHBs) are among the most important gravitational wave sources expected to be detected by the Laser Interferometer Space Antenna (LISA)~\cite{LISA:2017pwj}, an ESA mission in collaboration with NASA.
These systems are expected to be relatively rare, with \( \mathcal{O}(10) \) mergers per year, and located at cosmological distances, but their large masses produce extremely loud signals, with Signal to Noise Ratios (SNRs) of up to \( \mathcal{O}(1000) \)~\cite{LISA:2024hlh}.
Observations of MBHBs will provide new insights into the origin and evolution of supermassive black holes, as well as tests of general relativity in strong fields~\cite{LISA:2024hlh,LISA:2022yao}.
LISA is particularly sensitive to the mHz frequency band, which is precisely where these signals dominate.

Extracting the maximum amount of information from MBHB signals requires highly accurate --- and therefore computationally expensive --- waveform models, as well as efficient data analysis algorithms.
Traditional Bayesian methods, such as Markov Chain Monte Carlo (MCMC), become computationally expensive due to the high dimensionality of the parameter space that needs to be explored, and the costs of repeatedly evaluating the waveform models.

Most preliminary studies on simulated data have assumed stationary, Gaussian noise and well-behaved data to define a likelihood function \( p(\vec{x} | \vec{\theta}) \) for observed data \( \vec{x} \) and parameters \( \vec{\theta} \). 
In reality, these assumptions are only approximations and can introduce systematic errors~\cite{Burke:2025bun}.
Addressing these limitations may require adapting the likelihood function or treating the data more carefully, further increasing computational costs.

In this work, we take the first steps toward an alternative approach that leverages machine learning to replace the explicit likelihood with a fast, learned approximation trained on simulated data~\cite{Vilchez:2024qnw}.
Our current study focuses on isolated MBHBs in stationary Gaussian noise as a proof of concept.
Once this simplified scenario is well understood, these restrictions can be lifted by simply modifying the simulation pipeline without changing the inference algorithm itself.

\section{Sequential Neural Likelihood (SNL)} 

SNL is a simulation-based inference (SBI) method~\cite{Papamakarios:2018zoy}.
SBI methods replace a component of Bayes' theorem with a flexible parametric estimator, typically implemented using a neural network.
The network is trained on simulated data to approximate the chosen quantity, after which the model can be used for inference~\cite{Cranmer:2019eaq}.
Within the SBI framework, we choose Neural Likelihood Estimation (NLE), where the true likelihood \( p(\vec{x} | \vec{\theta}) \) is approximated by \( p_{\vec{\varphi}}(\vec{x} | \vec{\theta}) \), which we model with a normalizing flow parameterized by \( \vec{\varphi} \)~\cite{Rezende:2015ocs,Papamakarios:2017tec}.
Given a dataset of simulation-parameter pairs \( (\vec{x}, \vec{\theta} ) \) drawn from the joint distribution \( p(\vec{x}, \vec{\theta}) \) --- that is, by first drawing \( \vec{\theta} \) from a prior and running it through the simulator --- the training objective for SNL is to maximize the expected likelihood estimate:
\begin{equation}
    \vec{\varphi}_{\mathrm{opt}} = \argmax_{\vec{\varphi}} \mathbb{E}_{( \vec{x}, \vec{\theta}) \sim p(\vec{x}, \vec{\theta})} \left[ p_{\vec{\varphi}}(\vec{x} | \vec{\theta}) \right].
\end{equation}

The motivation for approximating the likelihood, rather than the posterior, is that the likelihood is prior-independent.
Sequential Neural Likelihood (SNL) takes advantage of this by iteratively refining the likelihood estimate for a specific observation \( \vec{x}_o \) over several rounds\cite{Papamakarios:2018zoy}.
Each round draws samples of \( \vec{\theta} \) from the current posterior estimate\footnote{For the first round, the posterior estimate is the prior \( p(\vec{\theta}) \).}, 
\begin{equation}
    p_{\vec{\varphi}}(\vec{\theta}|\vec{x}_o) \propto p_{\vec{\varphi}}( \vec{x}_o | \vec{\theta}) \, p(\vec{\theta}),
\end{equation}
runs the simulator for those samples, and appends the resulting pairs to the training dataset before resuming training.
This process is repeated until we reach convergence.

A key challenge is that the likelihood is defined as a distribution over all possible values of \( \vec{x} \), which, even for relatively short-lived gravitational wave signals, can have tens of thousands of dimensions.
Modelling such high-dimensional distributions is computationally infeasible, so we introduce a dimensionality reduction step in the data pipeline.
The degree of compression and associated information loss strongly influence both accuracy and convergence speed.

Specializing the model to a single observation means that SNL does not provide amortized inference for new detections, unlike some other SBI methods. 
However, focusing training on a shrinking region of parameter space enables a faster training process with fewer simulations and more modest hardware requirements.
Posterior sampling still requires MCMC, but this step does not involve additional waveform evaluations and is trivially parallelizable, making it far cheaper than full-likelihood approaches, especially as waveform complexity increases.

\section{Data Generation Pipeline} 
Our experiments rely on a fast simulation pipeline, designed to generate large volumes of MBHB coalescence data efficiently.
The main steps are outlined here, with full details in Secs.~4 and~5 of Ref.~\cite{Vilchez:2024qnw}.

The pipeline begins with waveform generation using the IMRPhenomD model~\cite{Husa:2015iqa,Khan:2015jqa} and the frequency-domain LISA response implemented in \texttt{lisabeta}~\cite{Marsat:2018oam,Marsat:2020rtl}.
The waveform model assumes aligned spins orthogonal to the orbital plane and includes only the dominant \( \ell = |m| = 2 \) mode.
The output consists of second-generation frequency-domain Time-Delay Interferometry (TDI) channels \( (\tilde{A}, \tilde{E}, \tilde{T}) \), but we discard \( \tilde{T} \) since its gravitational wave content is suppressed at low frequencies~\cite{Prince:2002hpa}.

These templates are whitened with reference LISA noise Power Spectral Densities (PSDs) \( S_n^{A, E}(f) \), enabling the addition of stationary Gaussian noise by sampling from a standard normal distribution.
The whitened signals are then transformed to the time domain via inverse Fast Fourier Transform.

The simplest approach to dimensionality reduction is Principal Component Analysis (PCA)~\cite{JolliffePCA}, representing each signal as the weights of its most informative components.
The top 128 components preserve more than 98\% of the variance, but the reconstruction quality is uneven: while the merger is nearly perfect, while the earlier inspiral is of lower quality.
This is due to the much lower amplitude of the inspiral compared to the merger.
Scaling the waveforms before PCA to equalize amplitudes introduces severe information loss overall, making this approach unsuitable.

Because PCA is inherently linear, we also investigated non-linear methods such as autoencoders.
These are neural networks with a bottleneck built into their architecture.
They are trained to reconstruct the original signal (or the noise-free signal in the case of noisy data).
If reconstruction is accurate, the bottleneck representation can serve as the compressed data.
Our autoencoders, trained on scaled data, reconstructed the inspiral and ringdown adequately but failed to capture the merger, likely because its short duration and high-frequency content were deemed unimportant during training.
This effectively introduces gaps at the most informative part of the signal, degrading inference performance.
While current results are limited by architecture choices and hardware constraints, future work will explore more powerful architectures and training strategies to improve reconstruction.

\section{First Experiments} 

\begin{figure}
    \centering
    \begin{subfigure}{.32\textwidth}
        \centering
        \includegraphics[width=0.99\linewidth]{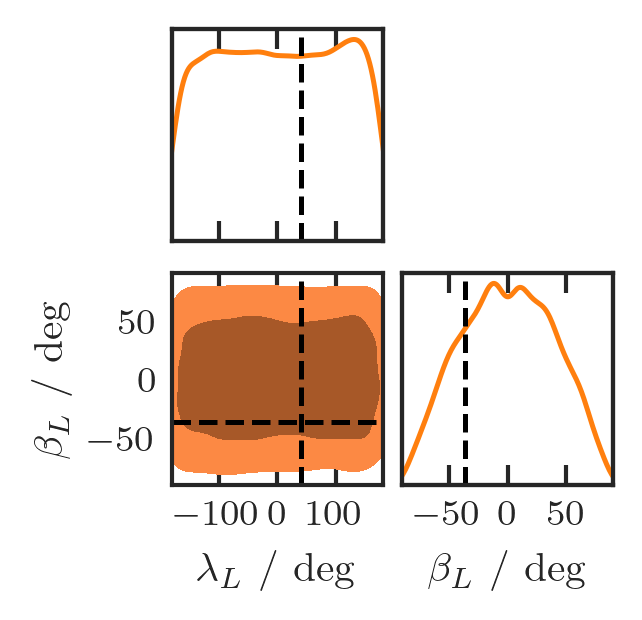}
        \caption{Round 1}
    \end{subfigure}
    \begin{subfigure}{.32\textwidth}
        \centering
        \includegraphics[width=0.99\linewidth]{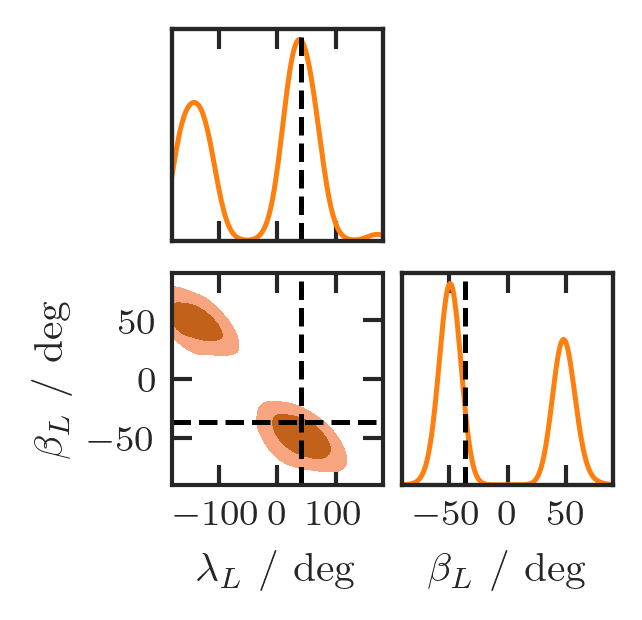}
        \caption{Round 15}
    \end{subfigure}
    \begin{subfigure}{.32\textwidth}
        \centering
        \includegraphics[width=0.99\linewidth]{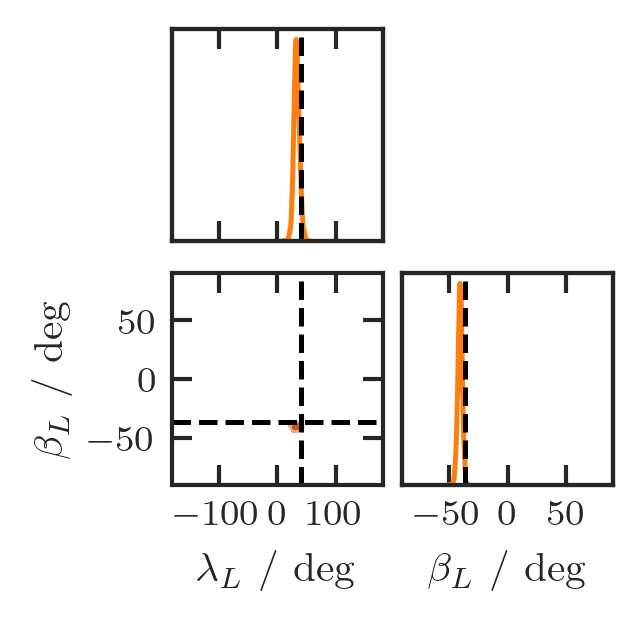}
        \caption{Round 90}
    \end{subfigure}
    \caption{Evolution of the sky location posteriors as SNL-PCA rounds on noise-free data progress. The diagonal subplots show the 1D marginalized posteriors for the longitude \( \lambda_L \) and latitude \( \beta_L \) in the LISA frame. The off-diagonal pane contains the 68\% and 95\% confidence regions for the 2D posterior. The dashed black lines indicate the true value for the parameters.}
    \label{fig:snl-evolution}
\end{figure}

We conducted experiments to assess the feasibility of SNL for MBHB parameter estimation.
Here, we just highlight some results and illustrate them with Fig.~\ref{fig:snl-evolution}. 
See~\cite{Vilchez:2024qnw} for a complete discussion.

We have tested three configurations: \textit{Eryn}, a likelihood-based parallel-tempered MCMC baseline, named after the sampler used~\cite{Karnesis:2023ras}; \textit{SNL-PCA}, SNL with PCA compression of unscaled data; and \textit{SNL-AE}, SNL with autoencoder-based compression of scaled data.
For SNL runs, training was stopped after 100 rounds, each adding \( 10^4 \) new simulations to the training dataset, for a total of \( 10^6 \) simulator calls --- less than 2\% of those performed in Eryn.
Training took \( \sim 80 \) hours, with most improvements occurring in the first 20 rounds; later rounds yielded only marginal gains.
Convergence is not always gradual: some parameters (e.g. longitude \( \lambda_L \)) exhibited sudden breakthroughs after many rounds.
The number of simulations per round also influenced convergence speed.

\paragraph{Posterior quality:}
SNL-PCA posteriors are qualitatively similar to those from Eryn, albeit slightly broader.
SNL-AE posteriors are significantly wider, as was expected from the more aggressive compression, but remain conservative and always encompass the true values of \( \vec{\theta} \).
When noise is added, posteriors slightly broaden and correlations between parameters weaken, though overall consistency between algorithms is generally preserved.
Notably, even with suboptimal summaries, SNL-AE accurately recovers the chirp mass (thanks to its strong imprint on the inspiral, which is preserved), and estimates the merger time within a time bin (a few bins in the noisy case) despite lacking merger information.

\paragraph{Degeneracies:}
With noisy data, SNL-PCA exhibits a bimodality in sky location, consistent with the symmetry in the LISA response for short-duration signals~\cite{Marsat:2020rtl}.
It can be justified by the loss of information in the early inspiral under the noise, which shortens the effective signal duration.
As can be seen in Fig.~\ref{fig:snl-evolution}, this effect also appears in intermediate rounds on noise-free data, but this secondary mode shrinks and eventually vanishes with further training.
SNL-AE shows an eight-fold degeneracy, combining this effect with the degeneracy appearing in the low-frequency approximation of the LISA response~\cite{Marsat:2020rtl}, caused by the loss of all the high-frequency information in the merger.
Both effects are expected to vanish when higher-mode waveforms are included in the simulation pipeline, which should also accelerate convergence.
Phase and polarization remain challenging for SNL methods, while Eryn recovers them with bimodal structures.

\section{Conclusions} 
We have introduced a novel approach to gravitational wave data analysis based on SBI.
Our method trains an approximate likelihood iteratively, adapting to specific observations by dynamically requesting simulations in regions of high posterior density.
This strategy greatly reduces the number of waveform evaluations required, which makes it particularly suitable for rare transient events where accurate waveform models are expensive --- MBHBs in LISA being a prime example.

Our current analysis uses simplified MBHB waveforms and stationary Gaussian noise with a fixed PSD.
However, all physical and instrumental assumptions are encapsulated in the simulation pipeline, so increasing realism should not require major changes to the inference algorithm.
In fact, additional information from more realistic waveforms may accelerate convergence.

The main outstanding challenge is the information loss introduced by dimensionality reduction.
High-quality low-dimensional representations of the data are essential to match the accuracy of traditional methods and further improve convergence speed.
Since we have full flexibility in data representation before and after compression, there is significant room for improvement in this area.

\section{Acknowledgements}
IMV and CFS are supported by contracts PID2019-106515GBI00 and PID2022-137674NB-I00 from MCIN/AEI/10.13039/501100011033 (Spanish Ministry
of Science and Innovation, MCIN) and 2017-SGR-1469 and 2021-SGR-01529 (AGAUR, Generalitat de Catalunya). IMV has been supported by FPI contract PRE2018-083616 funded by
– 34 –MCIN/AEI/10.13039/501100011033 (MCIN) and “ESF Investing in your future”, and by NextGenerationEU/PRTR (European Union). This work
has also been partially supported by the program Unidad de Excelencia María de Maeztu CEX2020-001058-M (MCIN).

\bibliography{main}

\end{document}